# High Resolution BIMA Observations
# of CO, HCN, and $^{13}$CO in NGC 1068

Tamara T. Helfer[1] and Leo Blitz[2]

Department of Astronomy, University of Maryland, College Park, MD 20742

Received ______________________;     accepted ____________________

---

[1]email:   thelfer@astro.umd.edu,   postal   address   Radio   Astronomy   Laboratory,   601
Campbell Hall, UC Berkeley, Berkeley, CA 94720

[2]email: blitz@astro.umd.edu

## ABSTRACT

We present high-resolution CO, HCN, and $^{13}$CO maps of the inner arcminute of NGC 1068. The maps were made by combining observations from the Berkeley-Illinois-Maryland Association (BIMA) interferometer with single-dish data from the NRAO 12 m telescope; the maps therefore do not suffer the usual interferometric problems associated with the lack of small spatial frequency visibilities. Several features appear in the CO map which have not previously been observed: (1) a firm detection of CO line emission from a compact region centered on the nucleus of the galaxy; (2) the spectrum of the unresolved nuclear emission shows a triplet velocity structure characteristic of kinematically independent regions; and (3) the detection of a molecular bar, the extent and position angle of which are in good agreement with the 2 $\mu$m stellar bar. As seen in previous high-resolution images, the most intense CO emission is nonnuclear at a typical distance of 15″ from the center of NGC 1068. The structure and kinematics of this emission imply that this gas is distributed along the inner spiral arms and not in a ring. The bar's kinematic influence on the molecular gas in the spiral arms is modest, with typical ordered noncircular motions of $\lesssim$ 30 km s$^{-1}$ in the plane of the galaxy. Interior to the spiral arms, the bar's influence is more dramatic, as reflected by the twisted isovelocity contours in the CO and HCN velocity fields. We derive a pattern speed for the bar of 150–170 km s$^{-1}$kpc$^{-1}$. The position angle of the bar, $\sim$ 63°, is close to that of the jet emanating from the nucleus, $\sim$ 33°.

The surface density of molecular gas within the central 100 pc radius of NGC 1068 is the same as that in the central 200 pc radius in the Milky Way to within the uncertainties. There is evidence for an $m = 1$ kinematic mode in NGC 1068; we find the kinematic center of rotation to be displaced from the radio continuum center by about 2.9″, or 200 pc. The HCN image, in agreement with recent results from other interferometers and in contrast to the CO map, shows a strong concentration of emission centered on the nucleus. The ratio of integrated intensities of the HCN emission to that of CO is about 0.6 and is the highest ratio measured in the central region of any galaxy. The $^{13}$CO emission follows the general distribution of the CO emission. The average CO/$^{13}$CO ratio of integrated intensities in the spiral arms is about 13; this is similar to what is observed in the centers of other nearby galaxies.

*Subject headings:* galaxies:individual (NGC 1068)—galaxies:ISM—galaxies:kinematics—galaxies:nuclei—galaxies:Seyfert—galaxies:starburst—interstellar:molecules

## 1. Introduction

One of the outstanding problems in studies of energetic phenomena in galaxies is understanding to what degree starburst and nuclear activity interact with each other – can a circumnuclear starburst provide gas and other matter as fuel for an active nucleus? can a starburst help to channel gas onto a monster AGN? Since large quantities of cold gas are commonly found in Seyferts (e.g. Rickard et al. 1977, the original detection of CO in NGC 1068, as well as surveys by Bieging et al. 1981, Blitz, Mathieu, & Bally 1986, Heckman et al. 1989, & Meixner et al. 1990) and since the molecular gas must be intimately involved with any starburst activity, it is clear that an understanding of the connection between star formation and nuclear activity is closely related to an understanding of the role of molecular gas in these sources.

NGC 1068 is the best-studied example of a close hybrid starburst/Seyfert galaxy; the bolometric luminosity of this source is roughly equally divided between its unresolved active galactic nucleus and its extended starburst disk (Telesco et al. 1984). The first high-resolution CO map of this galaxy (Myers & Scoville 1987) revealed rich structure in a $30''$ ($\approx 3$ kpc) diameter ring of emission, coincident with the extent of the starburst emission inferred from radio, infrared and optical observations (Wilson & Ulvestad 1982, Telesco et al. 1984, Telesco and Decher 1988, Atherton, Reay, & Taylor 1985). The starburst ring and the active nuclear emission appear to be linked by a 2.2 $\mu$m stellar bar, the linear dimension of which matches the diameter of the ring (Scoville et al. 1988, Thronson et al. 1989).

Driven by the desire to link conceptually the fueling of AGN with starburst activity, considerable effort has gone into trying to determine the amount, structure and kinematics of molecular gas in NGC 1068. In recent years, Planesas, Scoville, & Myers (1991) and Kaneko et al. (1992) have presented high-resolution CO maps (made with the Owens Valley Radio Observatory and with the Nobeyama Radio Observatory interferometers) that are striking for several reasons: (1) The maps depict rich sub-structure within the kpc-scale starburst "ring" of emission centered on the nucleus. Planesas et al. argued on morphological grounds that the structure of the CO emission was two inner spiral arms rather than a ring; Kaneko et al. instead distinguished four large, discrete complexes rather than either a ring or spiral structure. In this paper, we will present kinematic evidence that the CO traces the inner spiral arms of NGC 1068 (see § 3.3). (2) The maps show little (OVRO) or no (NRO) CO emission emanating from the compact nucleus. The OVRO detection is inferred indirectly from an excess of emission received in the upper sideband of their continuum correlator relative to that in the lower sideband. Our BIMA observation yields a firm detection of line emission from CO in the nucleus. (3) Despite the similarity of the instruments used for the observations, there are significant differences between the maps produced at OVRO and NRO – the details of the structure within the molecular ring appear quite different in the two maps, and the compact nuclear source detected at OVRO does not appear in the NRO map. These differences occur despite the similar sensitivities of the two observations and cannot be reconciled as an effect of the factor of 2 difference in resolution between the maps. In § 2.1 and 2.3, we discuss in detail how NGC 1068 presents special problems for interferometric observations because of its location on the Celestial Equator and because of its extended structure in CO.

While observations of CO have traditionally been used to map the distribution of molecular gas in galaxies, we began HCN observations of NGC 1068 in order to probe the dense molecular gas that might be more intimately associated with the nuclear star formation in this galaxy (Solomon, Downes, & Radford 1992, Helfer & Blitz 1993). HCN emission traces gas densities $\gtrsim 10^5$ cm$^{-3}$, which is about two orders of magnitude greater than the critical density required to excite CO. We note that Jackson et al. (1993) and Tacconi et al. (1994) have also recently presented interferometric maps of HCN in NGC 1068 made at NRO (Jackson et al.) and at the IRAM telescope at Plateau de Bure (Tacconi et al.).

In this paper, we present observations of CO, HCN, and $^{13}$CO in NGC 1068 made with the BIMA array. We will demonstrate the importance of including the zero-spacing data in order to determine the line ratios in this source, and we note that because we have used the same instrument with the same baseline spacings for these observations, we have minimized the systematic errors in the measurement of the line ratios. In the following sections, we present the BIMA maps along with a detailed discussion of the distribution, kinematics and the role of the molecular gas in the inner few kiloparsecs of this galaxy.

## 2. Observations and Data Reduction

### 2.1. Interferometric Observations

Synthesis observations of NGC 1068 were made with the newly expanded Berkeley-Illinois-Maryland Association (BIMA) millimeter array, located in Hat Creek, California. The BIMA array presently comprises six 6.1 m diameter dishes that are situated at various stations along a T-shaped runway which extends 305 m East-West and 183 m North-South. The four newest dishes have a measured surface accuracy of about 25 $\mu$m rms and 3 mm aperture efficiencies of 0.7-0.75. The receivers use cooled SIS and Schottky diode mixers and can be tuned from 85 to 115 GHz under computer control in about 3 minutes; measured double sideband receiver temperatures range from 50-150 K over the 3 mm window. The upper and lower sidebands of the first mixer are separated in the correlator by phase-switching the mm local oscillators. A second local oscillator at 1270 MHz is used to mix the IF to the 70-900 MHz band, which is sent back to the lab on underground cables. The spectrometer is a digital correlator with a continuum bandwidth of 800 MHz and 2048 complex spectral channels which are divided between the upper and lower sidebands.

The correlator is quite flexible and can easily be configured in different modes that allow from 6 kHz spectral resolution over 6 MHz bandwidth to 3 MHz resolution over 800 MHz bandwidth in each sideband. The signal entering the correlator can be split into up to eight windows per sideband; up to four of these windows can be positioned independently in the 800 MHz IF, and the bandwidths of each of the different windows may be set to varying values in octave multiples from 6.25 MHz to 200 MHz. A more complete description of the BIMA interferometer is given by Welch et al. (1995).

The NGC 1068 observations were carried out from November 1993 through February 1994 and included mapping the 3 mm J=1-0 transitions of CO (115.27 GHz), HCN (88.63 GHz), and $^{13}$CO (110.20 GHz). The pointing and phase center of the observations was $\alpha$(J2000) = $02^h 42^m 40.7^s$, $\delta$(J2000) = $-00°00'48.0''$ and the receivers were tuned to a velocity of $v_{LSR}$ = 1137 km s$^{-1}$. The HCN and CO observations were each obtained in three array configurations that included baseline coverage from 12.2 m to 238 m, while the $^{13}$CO observations included baseline coverage to 128 m only. The correlator was configured to achieve 1.56 MHz (5 km s$^{-1}$ at 100 GHz) spectral resolution over 300 MHz (900 km s$^{-1}$ at 100 GHz) bandwidth in each sideband; the line emission was centered in the upper sideband, while the lower sideband was used to detect continuum emission. The time dependences of the phase and amplitude antenna gains were calibrated using the nearby quasar 0238+166, which was observed every 25-30 minutes for a short integration. The spectral dependence of the gains was calibrated using a 30 minute integration on the strong quasar 3C454.3. Observations of Uranus served to calibrate the absolute flux scale, which is estimated to be accurate to about $\pm$ 30%. The typical system temperatures for the observations ranged from 250 – 700 K at 88 GHz and 110 GHz; at 115 GHz, the system temperatures were typically 600 – 1300 K for five of the antennas and up to 2000 K for the sixth antenna.

The data were reduced and analyzed on Sun workstations using the MIRIAD reduction package developed at the BIMA member universities (Sault, Teuben, & Wright 1995). Continuum emission was detected in the CO and HCN observations of NGC 1068 and was subtracted in the visibility plane by fitting zero-order polynomials to the real and imaginary parts of those channels in the passband that were free of line emission. Since the continuum emission is extended rather than pointlike, we checked the robustness of the continuum subtraction by comparing maps of the continuum made using the fits with the more sensitive maps made using the entire 300 MHz of the lower sideband; the two sets of maps were in reasonable agreement. The CO (HCN,$^{13}$CO) data were smoothed to 3.13 MHz (6.25 MHz, 6.25 MHz) resolution and gridded, Fourier transformed and CLEANed in the usual way; the visibilities were weighted by the noise variance to account for the differences in system temperatures and gains of the individual antennas. We were able to process the CO data further by performing an iterative routine of a phase self-calibration of the visibility data using model CLEAN components, then transforming the visibility data again and CLEANing to improve the model source for the self-calibration process. This technique is effective at removing antenna-based variations in the phase gains and helps to recover flux that gets lost to sidelobes of the synthesized beam. This turned out to be very important in the case of NGC 1068, where the CO emission is distributed in a complex, rich structure on many size scales: because NGC 1068 is close to the Celestial Equator, the $(u,v)$ tracks are nearly confined to constant values of $v$; this causes the synthesized beam to have very strong and numerous sidelobes to the north and south in the plane of the sky (Figure 1). Since the CO emission at any given velocity from NGC 1068 has real features that are situated north and south of each other (see ahead to the channel maps shown in § 3.3) at distances similar to the separation of some of the beam sidelobes, the sidelobes must be removed very carefully from the dirty maps. To ensure careful sidelobe subtraction, each channel map was CLEANed with a low gain of 0.01 to a cutoff level of twice the rms noise of the dirty channel maps.

## 2.2. Single-Dish Observations

The single-dish observations were made during 1994 June 12-15 using the 3 mm SIS receiver at the NRAO 12 m telescope on Kitt Peak, Arizona [3]. We observed orthogonal polarizations using two 256 channel filterbanks, each with a spectral resolution of 2 MHz per channel. The pointing was monitored every 1-2 hours with observations of Mars and the quasar 0238+166, both of which were within 17° of NGC 1068 throughout the observations; the rms pointing accuracy is estimated to be 5″. Observations of Mars were also used to set the absolute flux scale. Typical system temperatures were 200 – 300 K at 88 GHz and 110 GHz and 340 – 620 K at 115 GHz. At 3 mm, the telescope main beam efficiency $\eta_m$ and forward spillover and scattering efficiency $\eta_{fss}$ are 0.62 and 0.72. Since the data were in units of $T_R{}^*$, we multiplied by $\eta_{fss}/\eta_m = 1.16$ to convert to $T_{MB}$. We observed in position switching mode and read out individual scans every 6 minutes.

In order to minimize aliasing in the process of generating low spatial frequency visibilities from the single dish maps (see Wilner & Welch 1994), we mapped NGC 1068 at least to the extent of the BIMA primary beam (100″ at 115 GHz, 132″ at 89 GHz) using $\lambda/2D$ spacing along a two dimensional grid. For the CO observations, we mapped along a 7×7 grid of 22″ spacing; the HCN observations were spaced at 29″ along a 5×5 grid. In $^{13}$CO, we mapped a 3×3 grid with half beamwidth spacing of 28″. The spectra were processed using the COMB reduction package developed at AT&T Bell Laboratories. The data from the two polarizations were averaged together, and a linear fit to the emission-free portions of the baseline was applied to each spectrum. Spectra which required higher than a linear-order fit were typically rejected, though there were a few spectra where second-order fits to the baselines were made; these few cases included only spectra where the line emission was strong and could clearly be distinguished from the baseline. Finally, all the spectra from a given position were averaged together. The data cubes were then transferred to the MIRIAD package for further processing and combination with the BIMA data.

## 2.3.    The Zero-Spacing Problem and Data Combination

By analogy with most kinds of imaging telescopes, the linear resolution of an interferometer is limited by the length of the longest baseline in the array. However, interferometers suffer the additional limitation caused by the existence of a shortest projected baseline. The lack of coverage of small spatial frequencies in the visibility plane causes the synthesized beam of an interferometer to sit in a very broad and shallow negative bowl. This has relatively little effect on observations of compact sources, but causes the instrument to couple poorly to more extended structures (see the appendix to Wilner & Welch 1994 for a detailed description of the effects of missing small as well as large spatial coverage on synthesis images). In particular, Wilner & Welch comment that "while synthesis observations are often said to be 'sensitive' to structures on scales $\lambda/S_{min}$, where $S_{min}$ is the length of the shortest baseline, the central brightness recovered from a gaussian distribution characterized by FWHM $= \lambda/S_{min}$ is only about 3%"! The most insidious aspect of the missing flux problem is that even if the interferometer recovers a majority of the flux from a source, there is no way to predict how the missing flux distorts the true source distribution.

The most straightforward way of filling in the zero-spacing data is to measure the small spatial frequencies directly using a single-dish telescope. This technique is most effective when the diameter of the single dish is at least twice as large as that of the individual interferometer dishes; then there can be sufficient overlap between the spatial frequencies sampled by both instruments to ensure a reasonable match between the two data sets. The 12 m telescope is thus ideally suited to fill in the zero-spacing data for BIMA observations.

### 2.3.1.    Details of the Data Combination

We combined the single dish and interferometer data in the manner described by Vogel et al. (1984) in their appendix (see also Wilner & Welch 1994). First, we made a model source distribution by deconvolving the 12 m beam from the single dish maps. To simulate the source distribution that the interferometer measures, we then convolved the result with the primary beam of the BIMA dishes. The 12 m beam was assumed to be a circular gaussian of size 55″ (71″, 57″) at 115 GHz (89 GHz, 110 GHz) (Wilner & Welch 1994; Jewell 1994, private communication); the BIMA primary beam was assumed to be a circular gaussian of size 100″ (132″, 106″) at 115 GHz (89 GHz, 110 GHz) (Lugten 1994, private communication). We regridded the resulting model on a 64×64 grid with 4″ spacing to aid the FFT algorithm with power-of-two sampling. Next, we compared the relative calibration of the two data sets. The 12 m observations covered projected spacings to 4.6 k$\lambda$ (3.5 k$\lambda$, 4.4 k$\lambda$), while the BIMA observations measured projected spacings > 2.5 k$\lambda$ (1.4 k$\lambda$, 2.5 k$\lambda$). We generated model visibilities corresponding to points sampled at BIMA in the range 2.5-3.8 k$\lambda$ (1.4-2.9 k$\lambda$) for the CO (HCN) observations, since the response of the single dish is not as reliable at higher spacings. In order to achieve good relative calibration between the two data sets, we used a gain for the NRAO telescope of 34 Jy K$^{-1}$ (39 Jy K$^{-1}$, 35 Jy K$^{-1}$) at 115 GHz (89 GHz, 110 GHz). Finally, we generated a fully-sampled set of visibilities from the single-dish model source distribution by sampling at points randomly distributed in the $(u, v)$ plane in the range 0-3.8 k$\lambda$ (0-2.9 k$\lambda$, 0-3.6 k$\lambda$). The number of visibilities generated was chosen to match the sensitivity of the single-dish observations.

The final channel maps were made by taking the Fourier transform of the BIMA and model NRAO visibilities and CLEANing the resulting maps as described in § 2.1. The maps of integrated intensity were made by taking the zeroth moment of channel maps over channels with emission and clipping at the 1$\sigma$ level. The sensitivities of the channel maps and maps of integrated intensity are given in the figure captions below.

### 2.3.2.   *The Special Case of NGC 1068*

In summary, we note that there may be two reasons that the Planesas et al. (1991) and Kaneko et al. (1992) CO maps disagree. The first is that the two maps both lack low spatial frequencies that determine large structures in the maps. The second is caused by the special location of NGC 1068 on the Celestial Equator, which makes for strong and numerous sidelobes that must be removed very carefully from the channel maps (see Figure 1 and § 2.1).

## 3.   Results

The CO and HCN maps of integrated intensity of NGC 1068 are presented in Figures 2 and 3. As has been seen previously in published synthesis maps, most of the CO emission is nonnuclear, while the HCN emission is strongly concentrated near the nucleus. In the CO map, we also identify features that have not previously been observed. The most spectacular of these is the appearance of a molecular bar which provides a continuous molecular connection between the northern and southern inner spiral arms. We have also imaged CO line emission from a compact source at the nucleus of NGC 1068. We discuss the details of these observations in the following sections, and we present a comparison of the emission from these two molecular species. We also present the first interferometric images of $^{13}$CO in NGC 1068.

We assume a distance to NGC 1068 of 14 Mpc ($H_o = 75$ km s$^{-1}$ Mpc$^{-1}$) so that $1''$ = 68 pc. The velocities in this paper are LSR velocities, given according to the radio convention $v_{rad}/c = \Delta\lambda/\lambda_o$, where $\lambda_o$ is the wavelength in the rest frame of NGC 1068. These velocities differ from the optical convention $v_{opt}/c = \Delta\lambda/\lambda$, where $\lambda$ is the observed wavelength, by $v_{opt}$ - $v_{rad} \approx 4$ km s$^{-1}$.

### 3.1.   Nuclear Emission

As recent maps by Jackson et al. (1993) and Tacconi et al. (1994) have shown, the HCN map shown in Figure 3 exhibits strongly concentrated emission towards the nucleus of NGC 1068. This was a very surprising result given the apparent lack of CO emission from the same region (Myers & Scoville 1987, Planesas et al. 1991, Kaneko et al. 1992). However, an important point is that the HCN maps recovered essentially all of the single-dish flux measured at the IRAM 30 m telescope (Nguyen-Q-Rieu et al. 1992) and the NRAO 12 m telescope (Helfer & Blitz 1993), while the previous CO maps missed a critical contribution from the visibilities with low spatial frequencies. This information was recovered in the BIMA map by including the low spatial frequencies from the NRAO 12 m, as discussed in § 2.3.

As shown in the integrated intensity map in Figure 2, the CO emission peaks up strongly around the nucleus of NGC 1068. A spectrum of the position of peak emission is presented in Figure 4; the thin line shows the spectrum from the interferometer data alone, while the heavy line shows the spectrum with the single dish data incorporated into the interferometer data. There are three distinct kinematic components shown in the figures, centered at velocities of 980 km s$^{-1}$, 1130 km s$^{-1}$, and 1250 km s$^{-1}$. The strengths of the two outer components are unchanged with the addition of the single dish data in Figure 4, but the strength of the component at the systemic velocity increases by a factor of about two to 2 K in the $4.1'' \times 3.7''$ beam. (This effect is the natural result of the single dish data being selectively sensitive to emission near systemic velocity at the central position. At this position, there is no extended emission at the outlying velocities, so the interferometer already recovered all the flux in the outer velocity peaks.) The HCN spectrum at the center (Figure 5; for better S/N at the central position see Figure 2a in Tacconi et al. 1994) has velocity limits that correspond to the central velocity peak in the CO spectrum. The most straightforward interpretation of the complex CO spectrum is that the central velocity feature comes primarily from a compact nuclear source that is spatially coincident with the region where HCN is emitted. The outer CO velocity features are kinematically connected to the larger-scale emission in the map and are each found only on one side of the kinematic line of nodes (see §3.6.1). The kinematic morphology suggests that the components represent an extended disk or ring of emission outside the physical region where the central CO and HCN are concentrated.

## 3.2. Molecular Bar

The most dramatic feature of the BIMA CO observation of NGC 1068 is the detection of a molecular bar which extends some 25″ at a position angle of ∼ 63°. The extent of the molecular bar is in good agreement with that of the 2.2 μm stellar bar imaged by Scoville et al. (1988) and Thronson et al. (1989), though the infrared bar has a somewhat smaller position angle of 45°- 48° (see § 4.1.2). The molecular bar shows more complex internal structure than does the IR bar. In particular, in the inner ± 8″ from the nucleus, the northeastern arm of the bar appears to be extended along a different angle from the southwestern arm. There also appears to be a linear offset between the northeastern and southwestern extensions, which is commonly seen with dust lanes in bars (see § 4.1.2). The gas kinematics in the central 10″ shows a clear response to the bar (see the velocity field presented in § 3.6.1), and we will argue below (§ 3.6, § 4.1) that NGC 1068 shows many of the characteristics of a classic barred spiral galaxy.

## 3.3. Spiral Arms

The CO map shown in Figure 2 clearly demonstrates that most of the CO emission from the inner kpc of NGC 1068 is distributed in two inner spiral arms, rather than a ring. As is typical of barred spiral galaxies, the spiral arms in NGC 1068 originate at the two ends of the molecular/infrared bar. Emission from the arms is also seen in the HCN map in Figure 3, but at a much lower level. The spiral arms can be distinguished kinematically; Figure 6 shows a position-velocity diagram in CO along the major axis of the bar. The strong feature at (-15″, 1258 km s$^{-1}$) marks the beginning of the southern spiral arm, where it branches off the southwestern end of the bar. The kinematically distinct feature at (-27″, 1218 km s$^{-1}$) marks the end of the northern spiral arm as it wraps around to the west. If the CO emission were a true ring, rather than spiral arms, then these two features would be much more continuous in the position-velocity diagram. A similar distinction is apparent in the spiral arms on the eastern side of the galaxy.

The individual channel maps of CO emission shown in Figure 7 display the rich structure associated within the inner spiral arms of NGC 1068. In each ∼ 8 km s$^{-1}$ channel map, typically two to four major condensations can be identified; in some channels, upwards of half a dozen apparently independent structures are seen. Most of these condensations appear to be resolved; they have typical sizes of ∼ 500 pc and masses of ∼ $10^8$ M$_\odot$. These structures are much larger than Giant Molecular Clouds (GMCs) in the Milky Way, which have sizes of ∼ 100 pc and masses of $10^5$–$10^6$ M$_\odot$. Instead, they are similar to associations of Giant Molecular Clouds (GMAs) as defined in M51 by Vogel, Kulkarni, & Scoville (1988). We identified in our channel maps all the major condensations seen in the OVRO CO map as tabulated by Planesas et al. (1991) and found a reasonable agreement with their identifications. However, the line widths of the BIMA features are often larger, sometimes by a factor of two or more, than those tabulated by Planesas et al. (especially in the weaker associations in their table). This might be because the wings of the lines get lost in the noise in the OVRO spectra; in the more sensitive BIMA maps, the wings are detected and the resulting linewidths are broader. Planesas et al. noted that the masses derived from their CO line intensities are typically a factor of 2 or more lower than the derived virial masses for their associations, and they suggest that the discrepancy be resolved by reducing the adopted CO-to-$H_2$ conversion ratio by a factor of about four from that often assumed for the Milky Way. We note that the discrepancy may be resolved instead by including the full velocity extent of the line profiles in the virial mass determination. We also note the smooth progression of strong clumps of emission from channel to channel in Figure 7; the identification of a unique condensation in position and velocity is hampered by this progression. The linewidths of the individual associations are confused by the rotation of the galaxy and should probably not be trusted implicitly for virial analysis.

### 3.4.   HCN/CO Ratio

We convolved the CO map to the same beam size as that of the HCN map in order to compare directly the HCN and CO line intensities. A greyscale representation of the ratio of integrated intensities I(HCN)/I(CO) is presented in Figure 8. The ratio is typically 0.1 in the region of the spiral arms and increases roughly monotonically to 0.49 at the center of the map. Based on the assumption that the HCN nuclear emission comes from the same physical region as the central component of the the triplet structure in the nuclear CO spectrum, we also measured the ratio I(HCN)/I(CO) over the velocity extent of the central CO component alone (see Figure 9); this ratio rises to 0.57. The sharp increase in the ratio implies that the region of emission over the velocity limits of the central peak is physically distinct from the region that emits the outer velocity components, as argued in § 3.1; otherwise the ratio would remain the same regardless of the velocity limits of integration (as is the case in the region of the spiral arms in Figure 9).

The peak ratio of 0.57 is representative of the physical region corresponding to the linear extent of the synthesized beam, which is d ∼ 350 pc. To investigate the ratio on smaller scales, we made uniformly weighted HCN and CO maps; uniform weighting produces maps of smaller linear resolution at the price of lower sensitivity. The resulting maps had a linear resolution of $3.0'' \times 2.9''$, or d ∼ 200 pc. The peak ratio I(HCN)/I(CO) remained unchanged at 0.57 in the center of the map; this implies that the HCN and (central) CO are both emitted from the same physical region. We conclude further that either (a) the size of the physical region emitting the peak HCN and CO is small compared to the uniformly weighted beam of d ∼ 200 pc, or (b) the HCN and CO emission fill the naturally weighted beam, and the respective filling fractions of HCN and CO are the same over the size scales of the two sets of maps (though the filling fractions of the two molecular species need not be identical to each other).

## 3.5.    $^{13}$CO Results

Figures 10 and 11 show the integrated intensity and the velocity maps of $^{13}$CO emission in NGC 1068. The detected emission is confined to the spiral arms and to a first approximation follows the emission of CO in those regions. A detailed comparison of the $^{13}$CO and CO channel maps (Figures 11 and 7) reveals differences in the ratio of emission from these isotopes among the different associations of molecular clouds. This spatial variation in the I(CO)/I($^{13}$CO) ratio of integrated intensity is most likely caused by variations in the optical depth along the lines of sight to the different features. The ratio of measured line intensities I(CO)/I($^{13}$CO) range from 8 to 25 across the disk of NGC 1068, with an average of I(CO)/I($^{13}$CO) = 13 in the region of the spiral arms. This value is comparable to what is measured in the bulges and disks of other nearby galaxies (Rickard & Blitz 1985, Young & Sanders 1986, Sage & Isbell 1991).

We did not detect $^{13}$CO in the nuclear region of NGC 1068. The $2\sigma$ upper limit on the $^{13}$CO line temperature at the nucleus is 0.1 K in the $7.7'' \times 7.3''$ synthesized beam. For comparison, the peak CO line temperature at the nucleus is 0.6 K when convolved to the same large beam size.

## 3.6.    Kinematics

### 3.6.1.    Spiral Arms

The velocity field of the CO emission is shown in Figure 12. The field is well ordered in the region of the spiral arms and is in good agreement with that shown by Planesas et al. (1991) and Kaneko et al. (1992). To a first approximation, we see that the gas in the arms seems to be in simple rotation, with a kinematic major axis of about 90°. The closed contours to the west and east of the nucleus indicate that the rotation curve peaks at some radius $R_{max}$ and falls off at larger radii (as is seen in HI by Brinks et al. 1992). A closer inspection of the velocity field in the arms reveals that the velocity gradient across the northern part of the galaxy is more gradual than it is to the south. In the inner $10''$, the isophotes appear twisted in much the same way as seen in the H$\alpha$ velocity map shown by Cecil, Bland & Tully (1991); this twisting is the classic signature of the gas response to a bar potential.

We solved for the CO rotation curve by fitting a tilted-ring model to the data using the ROTCUR program written by Begeman (1989) and adapted to the dynamics package

NEMO (Teuben 1994); further processing of the kinematic modeling was done using the NEMO package. ROTCUR assumes circular orbits and performs a nonlinear least squares fit to up to six parameters as a function of ring radius; the fitted variables may include the systemic velocity, the coordinates of the center of rotation, the inclination, the kinematic line of nodes and the rotation velocity. We solved for the rotation curve by an iterative process, starting from the region of the spiral arms and working in towards smaller radii. Because the CO shows blended lines in the nucleus, the fits interior to about 5″ are not reliably determined. Based on our preliminary fits, we adopted a kinematic center of (+1.5″,-2.5″) relative to the radio continuum center and an inclination of 40°, a value in good agreement with those in the literature. Using fixed values for the central position and inclination, we then solved for the systemic velocity and adopted a value of 1130 km s$^{-1}$. The resulting rotation curve and kinematic position angle are shown in Figure 13; the average position angle over the region of the spiral arms (10″–24″ in Figure 13) is 84° ± 2°. The rotation curve does not achieve its maximum velocity right away, but rises gradually and linearly until it flattens at about r = 10″ (about 700 pc).

We modeled the two-dimensional velocity field using the derived rotation curve and position angle and compared the result to the CO data by subtracting the model from the data and examining the velocity residuals, which are shown in Figure 14. The residuals show an ordered spiral structure; this is indicative of noncircular motions and may be due to streaming motions along the spiral arms or to elliptical orbits influenced by the bar. The details of the structure in the velocity residuals depend to some extent on the exact choice of parameters used for the model velocity field; however, for all the models we ran, the velocity residuals were ordered. We also note the large residuals around the nuclear region of NGC 1068; these are caused in part by the confusion due to the complex CO line profiles at the center.

### 3.6.2.   Nucleus

Figure 15 shows an enlargement of the CO and HCN velocity fields in the vicinity of the nucleus. Again, the field is well ordered and to first approximation it looks as if the gas is in simple rotation; however, a crucial distinction between the nuclear and nonnuclear velocities is that the nuclear kinematic position angle is more like the position angle of the infrared bar (45°), rather than ∼ 84° as is the case for the spiral arms. This change in position angle suggests that the nucleus is kinematically distinct from the gas at larger radii. An inspection of Figure 12 shows that the isovelocity contours are already starting to twist in the outer edge of the bar, and they become aligned nearly parallel to it.

## 3.7.   Continuum Emission

The 85 GHz continuum map of the nuclear region of NGC 1068 is presented in Figure 16. There are two distinct emission peaks, one at $(+1'',0'')$ and the other to the northeast at $(+2'',+5'')$ from the peak of the radio continuum emission (Wilson & Ulvestad 1983). A similar elongation to the northeast was seen at 3 mm by Planesas et al. (1991) and by Tacconi (1994, private communication). The position of the northeast source is roughly the same as the peak of the 6 cm radio jet (Wilson & Ulvestad 1983). We measure a flux of 42 $\pm$ 8 mJy in the central continuum source and 41 $\pm$ 8 mJy in the northeast source.

## 4. Discussion

### 4.1. NGC 1068 as a Barred Spiral Galaxy

The CO and HCN data presented in this paper add to a growing list of observations that show that NGC 1068 exhibits all the characteristics of a classic barred spiral galaxy. First, there is the direct imaging of the bar in the infrared (Scoville et al. 1988, Thronson et al. 1989) and in CO (this paper). Second, while the velocity field is more or less regular at larger radii, the isovelocity contours at r < 8″ are dramatically twisted in the signature response to a barred potential. The kinematic line of nodes changes sharply from ∼ 84° in the region of the spiral arms to ∼ 135° in the circumnuclear region. These results are consistent with what is seen in the Hα velocity field (§ 4.1.1). Third, while the morphology of the CO bar is complex, it is very similar to what is seen in the dust lanes in barred spiral galaxies (§ 4.1.2). Finally, as we will argue below (§ 4.1.3), it may be that the optical light distribution interior to r < 15″ is strongly influenced by the presence of the bar. We review the kinematic and morphological evidence of the bar nature of NGC 1068 in the following subsections.

#### 4.1.1. Noncircular Motions in the Inner Disk

There is a contradictory history of evidence presented for both regular and noncircular motions in the inner disk of NGC 1068 that turns out to be intimately linked to the contradiction between the position angle of the optical light distribution (∼ 55°) and that of the kinematic line of nodes (∼ 85°). In general, the two cases can be stated as follows: (1) the kinematic line of nodes is forced to follow the optical light distribution at 55°, with the result that huge (∼ 100 km s$^{-1}$) noncircular motions are deduced, and (2) the kinematic line of nodes is a free parameter in the fit. In this latter case, the kinematic position angle is generally determined to be between 80°–90°, and the velocity field is well modeled by simple rotation. In the following paragraphs, we review the relevant literature, and we argue that the second case is the more natural one to adopt for understanding the global kinematics of the inner disk.

The position angle of the light distribution in the inner disk is about 55°, but the position angle of the isophotes gradually increases to about 90° beyond r > 30″ (seen in Sandage 1961, Hodge 1968 and others). Early optical studies of the kinematics of the inner disk of NGC 1068 were summarized and expanded on by Walker (1968), who found evidence for noncircular motions which he interpreted as expansion of the inner disk. Walker also found that motions at radii r > 30″ were well modeled by simple rotation, though (in contradiction with later studies) at a kinematic position angle of 55°.

More recently, Atherton et al. (1985) found in their Hα study that the kinematic position angle of rotation was about 90°. These authors reasoned that there must be substantial noncircular motions in the inner disk based on the difference in the kinematic and light distribution position angles; an inspection of their Hα velocity field (Figure 3 in their paper) reveals even more direct evidence for noncircular motions: the isovelocity contours in the inner 15″ appear twisted in a "Z"-shaped structure in the sense expected given the known orientation of the bar. (This twisting would have been quite difficult to interpret without *a priori* knowledge of the existence of the bar, especially given the blanking of their velocity field near the nucleus; indeed, these authors favor "possible large-scale expansion of the gas disc" to explain any noncircular motion.) The kinematic results of Atherton et al. were confirmed in the long-slit [OIII], Hβ study by Baldwin, Wilson & Whittle (1987), who determined that the rotation curve at radii > 15″ was flat and in simple rotation with a kinematic line of nodes at 80° ± 9°. Again, these authors infer the presence of noncircular motions interior to these radii, and indeed they argue that these motions are caused by "oval distortions or bar-like features with different position angles on different spatial scales." Finally, the kinematics of the narrow line region of NGC 1068 was examined in the impressive if ponderous [NII],Hα study of Cecil et al. (1990) (for a lighter read, try the conference proceedings of Cecil 1990). By applying a decomposition to the complex [NII] and Hα spectra, these authors show that the velocity field close to the nucleus and within ± 160 km s$^{-1}$ of systemic (see Figure 1 of Cecil or Figure 5 of Cecil et al.) is strongly influenced by the bar in the inner disk. The Hα velocity field of Cecil et al. is in excellent agreement with that presented in this paper.

The velocity field of the CO emission has thus far been determined by Myers & Scoville (1987), Kaneko et al. (1989), Planesas et al. (1991) and Kaneko et al. (1992). These data are all broadly consistent and show that the velocity field can be well represented by simple rotation with a position angle of 80°-90°. However, Myers & Scoville and Planesas et al. argue in favor of fixing the position angle at 55° to follow the optical light distribution on the grounds that the pitch angle for the two spiral arms then appears more continuous. Such a fit requires radial motions of ∼ 100 km s$^{-1}$ in the spiral arms, or about half the magnitude of the rotational velocity, and these authors argue that these large noncircular motions are caused by the influence of the bar.

It is worth noting however that both the magnitude and the pattern of the velocity residuals in the spiral arms are consistent with what one expects from spiral arm streaming if the bar and the disk have the different position angles discussed above. If we take the corotation radius of the bar to be near the ends of the bar (see §4.2), and assume that the pattern speed of the arms is the same as that of the bar (which seems reasonable considering that the arms emanate from the ends of the bar), then the pattern speed of the spiral arms will be rotating faster than the gas in the inner disk. Gas is accelerated as it approaches the spiral arms (and decelerated as it leaves); outside of corotation the acceleration induces a component to the velocity that is radially outward. If the southern half of the galaxy is the near side as implied by the HI absorption study of Gallimore et al. (1994), then the arms are trailing, as expected. For gas on the near side of the major axis, the outward radial motion is seen as a negative velocity residual along the length of the arm on the outside, and a positive velocity residual on the inside. The sense of the projected velocities is reversed on the other side of the major axis (the far side of disk), and one expects positive velocity residuals along the outside of the arm and negative velocity residuals inside.

This is precisely the situation that is pictured in Figure 14. The residuals are quite ordered along the length of the spiral arms, and the signs of the residuals are just what one expects from density-wave streaming. The magnitude of the expected velocity residuals depends on the size of the density perturbations of the arms, but values of $10-20\%$ of the rotation speed are common. Both the morphology and the velocity field of the CO are well represented as the response to a bar, with the inner spiral arms having a pattern speed similar to that of the bar.

Based on results from the literature and the BIMA CO measurements presented in this paper, the enormous radial motions required in the region of the spiral arms required by Myers & Scoville and Planesas et al. seem rather pathological. In the picture presented here, the bar's influence on the molecular gas in the region of the spiral arms is modest, driving a spiral pattern with typical noncircular motions of $\lesssim 20/\sin i$ km s$^{-1}$ (Figure 14). The influence on the molecular gas within about r < 8″ of the nucleus is more dramatic, with noncircular motions up to $\sim 80/\sin i$ km s$^{-1}$; here, the velocity field clearly is disturbed from simple rotation (Figure 12) in the same way as shown in the [NII],H$\alpha$ study of Cecil et al. (1990). The HCN velocity field at the nucleus (see Figure 15, also Jackson et al. 1993 and Tacconi et al. 1994) is consistent with that of the CO.

### 4.1.2. Morphology of the Bar

The original detection of the infrared bar (Scoville et al. 1988, Telesco et al. 1989) showed a regular and smooth light distribution such as is commonly observed in stellar light. By contrast, the CO distribution along the bar is very complex. In addition to the central concentration of CO, there are two "lanes" to the northeast and southwest (see Figure 2) that appear to point along slightly different position angles. The structure in the molecular gas is similar to the detailed structure often seen in dust lanes in barred galaxies of type SBb and later (Athanassoula 1984). Dust lanes are often found on both sides of the nucleus, but with a linear offset relative to each other; close to the nucleus, the dust lanes sometimes curl around the bulge. The modest curvature seen especially in the northeastern CO arm in NGC 1068 may be an example of Athanassoula's "curved" classification of dust lanes, where the concave side faces the major axis.

The CO map displays a number of characteristics seen in theoretical studies of the gas response to a barred potential (see Predergast 1983, Athanassoula 1992 and references therein). In models, the dust lanes are found near the leading edges of a bar. This is also the case for NGC 1068; if we assume that the southern half of the galaxy is the near side (as implied by the HI absorption study of Gallimore et al. 1994), then NGC 1068 is rotating in the counterclockwise sense, with the spiral arms trailing and the CO "lanes" on the leading edge of the bar. The dust lanes in models are often offset in position angle from that of the bar, as noted earlier for NGC 1068. Also, theory predicts that the bar region should be generally depleted of gas, except for concentrations at the center and at the ends of the bar, and except for enhancements along the shocks. Again, these features are all demonstrated in Figure 2.

We conclude that not only is the complex, lumpy structure in CO emission consistent with what is observed in other barred galaxies, but also that the morphology seen in CO is broadly consistent with the results of hydrodynamical modeling of gas flow in barred spiral galaxies.

It is worth pointing out that the position angle derived for the bar ($\sim 63°$) is similar to that of the jet emanating from the nucleus ($\sim 33°$) which seems to be impinging on dense interstellar gas (Wilson & Ulvestad 1987). The differences in position angle are significant, but the gaseous bar does seem to be able to provide the environment necessary to confine the jet to a relatively small distance from the nucleus.

### 4.1.3. Where is the Optical Bar?

Kinematics and morphology aside, if NGC 1068 is a barred spiral galaxy, why does it not appear barred in optical light? Thronson et al. (1989) suggested two possible answers:

first, that two different stellar populations are represented in the (infrared) bar and the (optical) disk; and second, that extinction plays a major role in obscuring the bar in optical light.

Figure 17 shows an archival *Hubble Space Telescope* image[4] of NGC 1068 taken with the WFPC using the wideband filter 439W. Although contaminated by extinction and possibly by the ionization cone, this image is very suggestive of the bar (A. Barth, private communication). There are clearly isophotes which are elongated along the same direction and which have the same linear extent as the bar. This elongated structure can also be seen in the ground-based photometric study by Schild, Tresch-Fienberg & Huchra (1985). As a possible reconciliation of the position angle controversy between the optical light distribution and the kinematic line of nodes, we suggest that the position angle of 55° that has traditionally been adopted for the optical light distribution in the inner 30″ of NGC 1068 is really no more than representative of the underlying bar potential, which then is obscured by dust as suggested by Thronson et al. (1989).

## 4.2. Resonances and the Nuclear Spectrum

We have computed the curves of $\Omega \pm \kappa/2$ for NGC 1068 based on the BIMA CO data and, to demonstrate the continuity to the outer disk, using the HI data of Brinks et al. (1995). (Here, $\Omega$ is the circular frequency and $\kappa$ is the epicyclic frequency, given by $\kappa^2 = Rd\Omega^2/dR + 4\Omega^2$; Binney & Tremaine 1987. We computed the derivative at each position by taking the average of the derivatives on either side of that point.) The $\Omega$ and $\Omega \pm \kappa/2$ curves are shown in Figure 18. We estimate the pattern speed of the bar $\Omega_b$ in several ways. First, we may take an upper limit to the pattern speed of the bar to correspond to the solid body portion of the rotation curve shown in Figure 13. A fit to the five innermost points gives a speed of 265 km s$^{-1}$ kpc$^{-1}$. Next, we note that the pattern of velocity residuals shown in Figure 14 and discussed in § 4.1.1 is consistent with gas streamlines lying between the corotation radius of the bar and the outer Lindblad resonance (OLR). This means that the OLR occurs at least as far out as the end of the bar, so that $\Omega_b$ is less than about 175 km s$^{-1}$ kpc$^{-1}$. Third, we note that the projected end of the bar is about 15″ from the nucleus in Figure 2; the deprojected distance is 1.3 kpc. If the corotation radius is at the end of bar (e.g. Binney & Tremaine 1987), then the rotation curve gives $\Omega_b$ at that radius, 160 km s$^{-1}$ kpc$^{-1}$. On the basis of numerical simulations of the gaseous response to a barred stellar potential, Athanassoula (1992) suggests that the corotation radius for bars lies at $1.2a \pm 0.2a$, where $a$ is the semimajor axis. This might push corotation out a bit farther, but it is unclear from our observations where one might expect $a$ of a stellar bar to lie with respect to the gas and whether the consequent 20% increase increase in the corotation radius is justified. In any event, it appears that the best estimate for $\Omega_b$ is between 150 – 170 km s$^{-1}$ kpc$^{-1}$.

---

[4]Based on observations made with the NASA/ESA *Hubble Space Telescope*, obtained from the data archive at the Space Telescope Science Institute, which is operated by the Association of Universities for Research in Astronomy, Inc., under NASA contract NAS 5-26555.

We can see from Figure 18 that the OLR then is very close to the distance where the interstellar medium in NGC 1068 turns from being primarily molecular to being primarily atomic. That is, essentially all of the molecular gas is found interior to radii r < 1.7 kpc, with the peak at r $\sim$ 1 – 1.4 kpc, and essentially all of the HI is found exterior to radii r > 2 kpc, with the peak of the inner HI ring at r $\sim$ 4 kpc (Brinks et al. 1994). The location of the OLR also corresponds to the region of the bright central portion of the galaxy associated with star formation seen in the Hubble atlas photographs.

In order for there to be an inner Lindblad resonance (ILR), $\Omega$ - $\kappa/2$ must exceed the pattern speed of the bar $\Omega_b$ at some location within corotation. In Figure 18, we see that $\Omega$ - $\kappa/2$ remains below $\Omega_b$ for all radii with the possible exception of the innermost point on the plot. Unfortunately, this point (indeed all those interior to about 5″) is not a reliable fit to the data, because of the blended spectral features at the center (§ 3.1). However, the data do not rule out the presence of an ILR, but rather constrain any ILR to be at a distance close to the location of the high density nuclear emission. On the other hand, the large velocity residuals at the center of the galaxy seen in Figure 14 suggest that large noncircular motions are present in the nucleus and that extrapolating inward may not be a valid procedure. It may in fact be the case that there is a bar within the bar at a scale of about $0.1a$, similar to that postulated by Shlossman et al. (1989) as a mechanism for fueling galaxies like NGC 1068. Such an inner bar would be a natural explanation of the large noncircular velocities very close to the nucleus. Establishing that these noncircular motions are indeed bar driven would require observations on an angular scale of about 1″ or less.

### 4.3.    A Possible $m = 1$ Mode in NGC 1068

Figure 19 shows the CO map with an overlay of ellipses which represent circles of constant radius in the plane of NGC 1068. The ellipses are centered on the peak of the CO light distribution (Table 1). The southeastern arm in NGC 1068 clearly appears to be closer to the nucleus compared to the northwestern arm. We recall that the kinematic modeling of CO in NGC 1068 (§ 3.6.1) yielded a kinematic center of rotation that was (+1.5″,-2.5″) relative to the center of the radio continuum emission (the pointing center of the BIMA map). Could these characteristics be indicative of an $m = 1$ mode in NGC 1068, i.e. an asymmetric contribution to the gravitational potential?

To investigate this possibility, we went back to the kinematic analysis described in § 3.6.1 in order to test the robustness of the determination of the kinematic center of NGC 1068. We fixed the kinematic center at $(0'',0'')$ and again solved for the rotation curve and position angle as a function of radius; for this test, we excluded emission at radii r < 8″ to minimize the complications from line blending and noncircular motions caused by the bar. The residual velocity field formed from the difference between this model and the CO data showed precisely the pattern expected if the (0,0) position is displaced from the true kinematic center (for a detailed description of velocity residuals characteristic of incorrect dynamical parameters, see van der Kruit & Allen 1978). We tried varying the offsets in right ascension and declination separately; each time, there resulted velocity residuals with a pattern indicative of a positional offset. It therefore appears that the 2.9″ (200 pc) displacement of the kinematic center from the position of the central radio continuum emission is a real effect. The asymmetry in the CO light distribution of the spiral arms, together with the offset between the kinematic center and the radio continuum center, are consistent with the existence of an $m = 1$ mode in the gravitational potential in NGC 1068.

## 4.4.    Physical Properties of the Center

### 4.4.1.    Comparison of Nuclear Ratio With Previous Observations

Though the nuclear ratio of I(HCN)/I(CO) $\approx 0.6$ is the highest measured in the center of any galaxy, previous reports listed this ratio as high as 1 or 2 (Jackson et al. 1993, Tacconi et al. 1994). A comparison of the HCN and CO results from all the interferometers is shown in Table 2 along with the $^{13}$CO results presented in this paper. This table lists the peak brightness temperature and the beam size of each of the observations. For direct comparison of the various observations, the effective peak temperature of a source of size $\theta_s = 2''$ is also given (based on Sternberg, Genzel, & Tacconi 1994); here T(2″) = T$_{peak}$ × $(\theta_B{}^2 + \theta_s{}^2)/\theta_s{}^2$, where $\theta_B$ is the synthesized beam size (or 4″, the total size of the nuclear emission source – whichever is bigger). The errors listed for T(2″) have been normalized to a channel width of 20 km s$^{-1}$; these are formal errors only, and the absolute line intensities are probably not accurate to better than ± 30% for any of the results quoted. The CO temperature of Planesas et al. (1991) is inferred from their continuum correlator and is not a direct measurement of a spectral line temperature (§ 1; see also footnote $c$ to Table 2). We note that the sensitivity listed by Kaneko et al. (1992) is probably underestimated, based on a direct comparison of channel maps and maps of integrated intensity shown in Planesas et al. (1991) and this paper. Their maps might have suffered from some of the problems that NGC 1068 is particularly sensitive to, as discussed in detail in § 2.

The nuclear ratio measured using BIMA is the most reliable to date for two reasons: First, these are the only interferometric maps that used the same instrument to measure both the HCN and CO – so not only is the $(u, v)$ coverage matched, but also the calibration and systematic errors are minimized. Second, the results presented here are the only measurements which incorporated zero-spacing data (§ 2.3). This latter point is very important for NGC 1068, since the CO peak temperature at the center then increased by a factor of two, while the HCN remained unchanged.

### 4.4.2. Nuclear Excitation

The ratio of integrated intensities of 0.6 measured over the central ∼ 200 pc of NGC 1068 is the highest measured in the nuclear region of any galaxy. Typical ratios for molecular gas found in the bulges of normal as well as starburst galaxies are 0.03–0.08 over kpc scales (Helfer & Blitz 1993 and references therein). For comparison, the ratio over ∼ 100 pc scales in solar neighborhood GMCs is ≲0.01 (Helfer & Blitz 1993, 1995a). Helfer & Blitz (1993) also showed that there is a tendency for the ratio I(HCN)/I(CO) to increase as a function of the linear extent over which the ratio is measured. This effect is seen in Figure 9, where the ratio I(HCN)/I(CO) increases by a factor of 5 from the kpc-scale starburst region to the central few hundred pc.

In a recent paper, Sternberg et al. (1994) argued that the unusually high HCN to CO ratio of integrated intensities might be explained in part by a selective depletion of gas-phase oxygen in the nuclear region of NGC 1068. As discussed in detail in Sternberg et al., there is evidence for such an underabundance of oxygen both from X-ray data (Marshall et al. 1993; Ueno et al. 1994) as well as from ultraviolet measurements (Snijders, Netzer & Boksenberg 1986; Kriss et al. 1992). Sternberg et al. calculate that, for oxygen abundances of less than about 0.4 times the solar abundance $[O]_\odot$, there results an unusually high relative abundance of HCN molecules relative to CO molecules, [HCN]/[CO]. While [HCN]/[CO] (*abundance* ratio, not ratio of integrated intensities) typically ranges from $10^{-4}$ in quiescent clouds in the Milky Way to $10^{-3}$ in hot cores in regions of active star formation, Sternberg et al. show that an oxygen abundance of $< 0.4 \, [O]_\odot$ results in abundance ratios [HCN]/[CO] $> 10^{-3}$. These authors ran large velocity gradient (LVG) models to quantify the physical conditions in the nuclear gas, and interpreted them in terms of an HCN to CO ratio of 2.

We also ran large velocity gradient (LVG) models, based on our measured 3 mm line intensities and the single dish HCN(4-3) and CO(4-3) measurements of Tacconi et al. (1994) (see Sternberg et al. 1994 for the tabulation of these measurements). There are two important assumptions that are made in applying this analysis: first, that the line emission from the various transitions of the same molecule are emitted from the same physical region, and second, that the filling fractions of the emission from the different transitions are the same. Given the factor of three difference in the beam sizes of the 3 mm and the submillimeter observations, the first assumption is certainly questionable. Since the HCN emission towards the nucleus is relatively compact and since the peak emission of HCN is dominated by the nuclear component, this assumption is probably reasonable for the HCN analysis. However, the case for the CO emission is not so clear. The larger beam sizes of the single dish transitions may pick up a significant contribution from the molecular bar, which is inside the half-power beam width of the submillimeter observations, and it may also include a significant contribution from emission from the spiral arms. Even on very small scales, the triplet structure in the BIMA central CO pixel implies that the CO emission within $2''$ of the nucleus comes from at least two physically distinct regions. We therefore treat the results of this analysis with caution: we interpret the results as guidelines to the physical conditions near the nucleus, and we award relatively higher weight to the HCN analysis.

Figure 20a shows the results of our HCN analysis. We assumed a kinetic temperature of $T_K = 50$ K and ran models for a range of molecular hydrogen densities and HCN column densities. As in Sternberg et al. (1994), we corrected for the effects of the different beam sizes by "deconvolving" the peak brightness temperature to that for a nuclear source size with an effective size of $\theta_s = 2''$ (§ 4.3.1, Table 2). The peak HCN(1-0) equivalent source temperature was $T_s = 5$ K, which according to Figure 20a constrains the HCN column density to be $N_{HCN} \gtrsim 1 \times 10^{15}$ cm$^{-2}$. Using the HCN(4-3) data from Sternberg et al., the ratio $T_s(4\text{-}3)/\ T_s(1\text{-}0) \sim 0.7$ then implies a density of molecular hydrogen $n(H_2) \sim 4 \times 10^6$ cm$^{-3}$ and $N_{HCN} \approx 2 \times 10^{15}$ cm$^{-2}$.

The results for the CO analysis are shown in Figure 20b. The density is less well constrained than with the HCN analysis, with a lower limit of $n(H_2) > 2 \times 10^4$ cm$^{-3}$ implied. The column density of CO is constrained to be $N_{CO} \approx 2\text{--}5 \times 10^{18}$ cm$^{-2}$. With $N_{HCN} \approx 2 \times 10^{15}$ cm$^{-2}$ and $N_{CO} \approx 2\text{--}5 \times 10^{18}$ cm$^{-2}$, we infer an abundance ratio [HCN]/[CO] $\approx 4\text{--}10 \times 10^{-4}$, or a value similar to that measured for Milky Way molecular clouds. While the effect described by Sternberg et al. might contribute modestly to the observed line ratios, we conclude that it is not required; excitation effects alone can explain the observed line ratios.

Our result differs from that of Sternberg et al. largely as the result of somewhat lower HCN and and somewhat higher CO peak temperatures, and because we have a lower observed ratio of HCN to CO peak temperature. Sternberg et al. had to rely on observations taken at three different telescopes. For their interferometric results, there are the additional problems that the temperature of the CO line was inferred from continuum observations, and the line strengths were measured by two instruments with different antenna spacings and without the inclusion of the zero-spacing data. The discussion in §2.3 summarizes the difficulties that this particular galaxy poses for interferometric observations. Although the absolute calibration of millimeter-wave sources continues to be somewhat problematical, our observations of both HCN and CO were done the same way and included the zero spacing data. We therefore believe that our line ratios are reliable.

### 4.4.3.   Comparison With the Milky Way

How does the surface density of molecular gas in NGC 1068 compare with that in the Milky Way? We can estimate the column density in the central r = 130 pc (the synthesized beam size) of NGC 1068 in two ways: first, we measure an integrated intensity of 350 K km s$^{-1}$ over the central spectrum shown in Figure 4b. Using a CO/H$_2$ conversion ratio $X$ = 2.5 × 10$^{20}$ cm$^{-2}$(K km s$^{-1}$)$^{-1}$, this implies N(H$_2$) = 8.8 × 10$^{22}$ cm$^{-2}$. In addition, the results of our LVG modeling above suggest that N(CO) = 2–5 × 10$^{18}$ cm$^{-2}$; if we assume a CO/H$_2$ abundance ratio of 8 × 10$^{-5}$, this implies a column density of N(H$_2$) = 2.5–6.3 × 10$^{22}$ cm$^{-2}$. We therefore adopt a column density of N(H$_2$) = 5 × 10$^{22}$ cm$^{-2}$; the surface density in the inner r = 130 pc of NGC 1068 is then about $\sigma_{\mathrm{NGC1068}}$ = 600 M$_\odot$ pc$^{-2}$. In the Milky Way, there is about 1 × 10$^8$ M$_\odot$ of molecular gas within the central r = 200 pc (Sanders, Solomon, & Scoville 1984, Sodroski et al. 1994); the surface density is thus about $\sigma_{\mathrm{MW}}$ = 800 M$_\odot$ pc$^{-2}$. Although there is a factor of 2–3 uncertainty in the determinations of both $\sigma_{\mathrm{NGC1068}}$ and $\sigma_{\mathrm{MW}}$ (mostly due to the uncertainty in $X$), the agreement between the central 130 pc of NGC 1068 and the central 200 pc of the Milky Way is remarkable.

Furthermore, the offset of the kinematic center from the nucleus in NGC 1068 is similar to the offset that is observed of the central molecular gas disk from the assumed dynamical center at Sgr A* (e.g. Blitz et al. 1993). In the case of the Milky Way, the projected offset is about 80 pc, and there is an offset in the central velocity as well of about 32 km s$^{-1}$. Evidently, the $m = 1$ mode that is inferred from our observations of NGC 1068 is not unique either in its existence or its magnitude, though its origin, as is true for the $m = 1$ mode in the Milky Way, is unclear.

An important distinction in the distribution of molecular gas in NGC 1068 and the Milky Way is that in NGC 1068, there are massive molecular arms close to the center (r = 0.6–2 kpc) which the Milky Way does not have. In fact, there is a relative paucity of molecular gas in the Milky Way at radii between r = 0.3 kpc and r = 3.5 kpc. The large reservoir of molecular gas in the inner spiral arms of NGC 1068 might be the source of fuel for the AGN. While the Milky Way has a similar concentration of molecular gas at its center and the same potential mechanism for channeling gas inward from larger radii (i.e. bar streaming), there is not a similar massive reservoir of molecular gas at kpc scales.

### 4.5.   $^{13}$CO Results

We ran an LVG analysis of $^{13}$CO based on our measured upper limit in the nuclear region of NGC 1068 and the $^{13}$CO J=3-2 single-dish measurement tabulated in Sternberg et al. (1994). If we assume an intrinsic source size of 2″, then the corresponding 3$\sigma$ upper limit to the effective peak $^{13}$CO temperature at the nucleus is T(2″) < 3 K. We find that the LVG analysis constrains the physical conditions in the nuclear region much more loosely than the HCN and CO analysis presented above: the $^{13}$CO data imply that the density is $> 10^3$ cm$^{-3}$ and the column density of $^{13}$CO is $< 10^{18}$ cm$^{-2}$. Assuming a $^{13}$CO/CO isotopic ratio of 1/40, this implies that $N_{CO} < 4 \times 10^{19}$ cm$^{-2}$. Since these do not constrain the physical conditions at the nucleus, we may ask, at what level could a $^{13}$CO measurement provide a useful constraint? If we assume the conditions inferred above from the HCN and CO analysis, then the nuclear $^{13}$CO source temperature should be about $T(2'') \approx 0.1\text{-}0.2$ K. Considering the typical errors quoted in Table 2, this would be a difficult measurement to make given the collecting areas of current millimeter interferometers.

In the spiral arms, the average ratio of I(CO)/I($^{13}$CO) is 13, with values ranging from 8 to 25 across the disk. These values are similar to what is observed in the centers of other nearby galaxies (Rickard & Blitz 1985, Young & Sanders 1986, Sage & Isbell 1991) and suggest that the physical conditions of the gas in the inner spiral arms are rather normal, in contrast to the extreme physical conditions in the high-pressure nuclear region (§ 4.4.2).

### 4.6.    What is Not Unusual about NGC 1068

While NGC 1068 has the highest *ratio* of HCN to CO emission measured over any galactic nucleus, it should be emphasized that NGC 1068 is *not* unusual for the existence or central concentration of HCN. Single-dish spectroscopic surveys (Mauersberger et al. 1989, Sage, Solomon, & Shore 1990, Nguyen-Q-Rieu et al. 1992, Israel 1992, Helfer & Blitz 1993) have demonstrated that most spiral galaxies, not just starburst galaxies, have an appreciable amount of dense gas in their bulges. The central $\sim 250$ pc of the Milky Way exhibits strong, extended emission by CS (Bally et al. 1987) and HCN (Jackson, Heyer, & Paglione 1993). We have recently imaged the HCN emission from four nearby spirals (NGC 3628, NGC 4826, NGC 5236, and NGC 6946) with the BIMA interferometer, and we find that as with the Milky Way and NGC 1068, the emission from dense gas is strongly concentrated within the central few hundred pc of these sources (Helfer & Blitz 1995b). What seems apparent from these extragalactic studies as well as our large-scale study of the disk of the Milky Way (Helfer & Blitz 1995a) is that there is a much greater similarity among the bulges of galaxies than there is from bulge to disk in a galaxy: dense gas is commonly observed over large ($\gtrsim 100$ pc) scales in the bulges of galaxies, while it is less widely distributed in disk molecular clouds. With $n_H \gtrsim 10^5$ cm$^{-3}$ and d $\gtrsim 20$ pc (the lower limit is set by the scale height of the molecular gas), column densities of order $N \sim 10^{23-24}$ cm$^{-2}$ are fairly easy to achieve towards the nuclei of (presumably) most spiral galaxies, even with a filling factor of less than unity. This result suggests that it is not necessary to invoke strange geometries or physical conditions to explain the high column densities derived from the hard X-ray spectra of Seyfert nuclei (e.g. Mulchaey, Mushotzky, & Weaver 1992).

## 5. Conclusions

We have presented high-resolution CO, HCN, and $^{13}$CO maps of the Seyfert/starburst hybrid galaxy NGC 1068 made by combining BIMA observations with single-dish NRAO data. We discussed how the extended structure of NGC 1068 and its special location on the Celestial Equator make it a particularly challenging source to image with an interferometer; these factors may help to explain the differences in the published CO maps of NGC 1068.

The major results of this study are as follows:

(1) We have detected a molecular bar which lies along roughly the same position angle and which has a similar linear extent as the 2 $\mu$m stellar bar. The kinematics of the molecular gas is clearly responding to the bar potential. In the spiral arms, the response is modest, with noncircular motions of $\lesssim$ 30 km s$^{-1}$ in the plane of the galaxy. Interior to the bar, the response is much more dramatic, as is reflected in the CO and HCN velocity fields. Within the bar, the CO exhibits structure similar to what is commonly seen in dust lanes in barred galaxies of type SBb and later. The CO "lanes" are found on the leading edge of the bar, and they appear to curl around the central emission. Except for the CO "lanes" and the concentrations at the center and ends of the bar, the bar region is generally depleted of molecular gas. Based on kinematic and morphological results presented in this paper and from the literature, we argue that NGC 1068 exhibits all the major characteristics of a classic barred spiral galaxy.

(2) We have imaged CO as well as HCN in the nuclear region of NGC 1068. The spectrum of the nuclear CO feature shows an unusual triplet structure, with velocity components at 980 km s$^{-1}$, 1130 km s$^{-1}$, and 1250 km s$^{-1}$. The HCN spectrum at the center has velocity limits that correspond to the central velocity peak in the CO spectrum. The most straightforward explanation of the complex CO spectrum is that the central velocity component is associated with the same compact nuclear source which emits the HCN; the outer CO velocity features may represent a ring or disk of emission outside the physical region where the nuclear molecular emission is concentrated. To within the uncertainties, the surface density of gas within the central 100 pc radius in NGC 1068 is the same as that within the central 200 pc radius in the Milky Way.

(3) In agreement with published results, most of the CO emission from NGC 1068 is nonnuclear. The morphology and kinematics of this region imply that the CO traces the inner spiral arms of NGC 1068 and not a ring. As is typical of barred spiral galaxies, the spiral arms originate at the ends of the molecular/infrared bar. The individual condensations within the spiral arms have typical sizes of $\sim$ 500 pc and masses of $\sim$ $10^8$ M$_\odot$; these are much larger than GMCs in the Milky Way and are probably instead associations of GMCs (GMAs).

(4) We measured the ratio of integrated intensities I(HCN)/I(CO) as a function of

radius in NGC 1068. The ratio is about 0.1 in the region of the spiral arms and increases monotonically to about 0.5 at the nucleus. If this ratio is taken over the limits of the central velocity feature in the CO spectrum, the ratio increases to 0.6. The nuclear ratio of 0.6 is the highest measured in the center of any galaxy. The physical conditions in the molecular gas at the center are well modeled by 50 K gas at a density of several $\times 10^6$ cm$^{-3}$.

(5) We presented the first interferometric maps of $^{13}$CO in NGC 1068. The I(CO)/I($^{13}$CO) ratio in the spiral arms is 8–25, with an average value of 13. These ratios are similar to what is seen in the centers and disks of other galaxies. We did not detect $^{13}$CO from the nucleus of NGC 1068, and the measured upper limit does not put appreciable constraints on the physical conditions in this region. The peak $^{13}$CO temperature at the nucleus suggested by our CO and HCN modeling is more than an order of magnitude lower than our measured upper limit; a temperature of this magnitude would be difficult to measure with the current millimeter interferometers.

(6) The southeastern CO arm is clearly closer to the nucleus of NGC 1068 than is the northwestern arm, and the kinematic center of rotation is displaced from the radio continuum center by about 2.9″ (200 pc). These features suggest an $m = 1$ mode in the gravitational potential in NGC 1068.

(7) Even though NGC 1068 has the highest ratio of HCN to CO emission measured in any galaxy, NGC 1068 is not unusual for the existence or central concentration of HCN. Most spiral galaxies have an appreciable amount of dense gas in their centers, and recent imaging of HCN in four nearby spiral galaxies with the BIMA interferometer (Helfer & Blitz 1995b) show that dense gas is strongly concentrated within the central few hundred pc of the nuclei in these sources. These results suggest that high column densities are fairly easy to achieve in most spiral galaxies, and that it is not necessary to invoke strange geometries or physical conditions to explain the high column densities derived from the hard X-ray spectra of Seyfert nuclei.

Many engineers and scientists have worked hard to build, operate and improve the expanded BIMA array; this work would not have been possible without their labor. We thank Phil Jewell and the NRAO 12 m staff (especially telescope operators Lisa Engel and Harry Stahl) for assistance with the single dish observations. We thank Jack Gallimore, Andy Harris, Tom Hartquist, Jim Jackson, George Miley, Bikram Phookun, Glenn Piner, Mike Regan, Frank Shu, Amiel Sternberg, Jim Stone, Linda Tacconi, Peter Teuben, and Mel Wright for helpful discussions. Peter Teuben gave considerable assistance with the kinematic modeling, and Lee Mundy provided the LVG code. Jack Welch generously provided an office, computer and other amenities for TTH to use while visiting UC-Berkeley. LB would like to thank the Leiden Observatory for its hospitality during the last stages of the writing of this paper. Special thanks go to John Lugten for many conversations and suggestions on the details of the data reduction and analysis. We thank Aaron Barth for pointing out the suggestion of the bar in the HST image and for retrieving the HST image from the STScI data archive. Eli Brinks very kindly sent us the HI data in advance of publication. This work was partially supported by a grant from the National Science Foundation, with additional support by a grant from the State of Maryland.

Table 1.   Centroid Positions in NGC 1068

| Centroid | $\Delta\alpha, \Delta\delta$ |
|---|---|
| 6 cm radio continuum[a] | $0'', 0''$ |
| 3 mm continuum | +1, 0 |
| CO light distribution | 0, -1 |
| HCN light distribution | +1, -1 |
| Kinematic center | +1.5, -2.5 |

[a] Wilson & Ulvestad 1983

Table 2: Interferometric Measurements of HCN, CO and $^{13}$CO
in the Nucleus of NGC 1068

| Reference | $T_{peak}$ (K) | Beam ($'' \times ''$) | $T(2'')^a$ (K) |
|-----------|----------------|----------------------|----------------|
| HCN | | | |
| Jackson et al. (1993) | 0.58 | $7.4 \times 6.2$ | $7.2 \pm 1.6^b$ |
| Tacconi et al. (1994) | 1.7 | $5.0 \times 5.0$ | $12 \pm .09$ |
| This paper | 0.61 | $5.4 \times 4.8$ | $4.6 \pm 0.8$ |
| CO | | | |
| Planesas et al. (1991) | $1.2^c$ | $2.9 \times 2.9$ | $6.0^c \pm 2.6$ |
| Kaneko et al. (1992) | $< 0.1^d$ | $5.4 \times 4.2$ | $< 1.1^d$ |
| This paper | $2.0^e$ | $4.1 \times 3.7$ | $10^e \pm 1.3$ |
| $^{13}$CO | | | |
| This paper | $< 0.1$ | $7.7 \times 7.3$ | $< 2$ |

$^a$ Peak temperature normalized to source size of $2''$ (see text) .

$^b$ Errors normalized to 20 km s$^{-1}$ channel width for direct comparison.

$^c$ Inferred from continuum measurement as
$T_{peak} = 19$ mJy $\times$ 1040 km s$^{-1}$ $\times$ 10.9 K Jy$^{-1}$ / 175 km s$^{-1}$.

$^d$ $2\sigma$ upper limits; error is probably underestimated (see text).

$^e$ From Figure 4b; a gaussian decomposition to the central velocity component yields
$T_{peak} = 1.5$ K and $T(2'') = 7.5$ K.

# FIGURE CAPTIONS

**Figure 1** (*a*)BIMA *uv* coverage for the NGC 1068 CO observations. Since NGC 1068 lies very nearly on the Celestial Equator, the coverage is confined to nearly constant values of *v*. This gives rise to (*b*) a synthesized beam with strong and numerous sidelobes to the north and south. The contours are ± 10,30,50,70, and 90% of the peak beam response. For comparison, the BIMA *uv* coverage and synthesized beam for a similar observation of the phase calibrator, which lies at $\delta = 16°$, are shown in (*c*) and (*d*).

**Figure 2** BIMA integrated intensity map of CO in NGC 1068. The naturally weighted beam is 4.1″ × 3.7″ and is shown in the lower left corner. The contours are shown at ± 4,6,7,8,10,12,14 $\sigma$ levels, where $\sigma = 5.7$ Jy beam$^{-1}$ km s$^{-1}$. The 7$\sigma$ contour is included to highlight some of the emission from the molecular bar.

**Figure 3** BIMA integrated intensity map of HCN in NGC 1068. The naturally weighted beam is 5.4″ × 4.7″ and is shown in the lower left corner. The lowest level contour is ± 3$\sigma$, and the contour interval is 2$\sigma$. The sensitivity in the map is $\sigma = 1.5$ Jy beam$^{-1}$ km s$^{-1}$. The sensitivity in an individual channel map of 21 km s$^{-1}$ (not shown) is 0.03 Jy beam$^{-1}$.

**Figure 4** Spectra from the central 4.1″ × 3.7″ beam of the NGC 1068 CO map. (*a*) The thin line shows the spectrum from the interferometer data alone, while (*b*) the heavy line shows the spectrum from the combined BIMA/NRAO map, which includes the zero-spacing data. (*c*) The combined spectrum is shown as an overlay on the spectrum from the interferometer data.

**Figure 5** Spectrum from the central 5.4″ × 4.7″ beam of the NGC 1068 HCN map. The thin line shows the spectrum from the interferometer data alone, while the heavy line shows the spectrum from the combined BIMA/NRAO map.

**Figure 6** Position-velocity cut along the major axis of the CO bar. Contours are in ± 1.5$\sigma$ steps from 2$\sigma$ at the lowest level, where $\sigma = 0.07$ Jy beam$^{-1}$. Distances are measured along the position angle (63°) of the bar axis from the kinematic center of rotation (see § 3.6.1). Positive distances represent positions to the northeast of the kinematic center of rotation.

**Figure 7** Velocity maps of CO emission; each map represents an 8 km s$^{-1}$ channel. LSR velocities are shown in the upper left corner and the synthesized beam is shown in the lower left corner of each channel map. The contour interval is 2$\sigma$, with the lowest contour at 3$\sigma$. The sensitivity in an individual 8 km s$^{-1}$ map is $\sigma = .07$ Jy beam$^{-1}$.

**Figure 8** Greyscale ratio map of integrated intensities I(HCN)/I(CO) in NGC 1068. The CO map was convolved to the same beam size as that of the HCN, and the integrated intensities were converted to $\int T_{MB}$km s$^{-1}$ (K km s$^{-1}$) before the ratio was taken. The halftone limits are (0,0.5).

**Figure 9** Graph of the ratio of integrated intensities I(HCN)/I(CO) as a function of radius

in NGC 1068. The boxes represent the ratio of integrated intensities over the entire range of velocities represented in the maps. The crosses represent the ratio over the velocity limits of central CO feature only (see text). The error bars shown are based on the number of points included in each ring of the elliptical integration and are formal errors only; the measurement uncertainty in the peak ratio of I(HCN)/I(CO) is $\pm$ 0.1.

**Figure 10** BIMA integrated intensity map of $^{13}$CO in NGC 1068. The naturally weighted beam is $7.7'' \times 7.3''$ and is shown in the lower left corner. The lowest level contour is $\pm 2\sigma$, and the contour interval is $1\sigma$. The sensitivity in the map is $\sigma = 2.4$ Jy beam$^{-1}$ km s$^{-1}$.

**Figure 11** Velocity maps of $^{13}$CO emission. Each map represents a 17 km s$^{-1}$ channel, and the LSR velocities are given in the upper left corners. The contour interval is $1\sigma$ with the lowest level at $\pm 2\sigma$, where $\sigma = 0.04$ Jy beam$^{-1}$. The naturally weighted synthesized beam, shown in the lower left corner of each map, is $7.7'' \times 7.3''$.

**Figure 12** The CO velocity field is shown as contours overlaid on the CO intensity halftone. Contours are shown in 25 km s$^{-1}$ steps over the range 900–1350 km s$^{-1}$. The systemic velocity contour of 1130 km s$^{-1}$ is shown as a heavy line. Smaller velocities are to the east of systemic, and the velocities increase to the west.

**Figure 13** Results from the ROTCUR kinematic modeling of CO in NGC 1068. The error bars are based both on the number of points used in the fit and on the reliability of the least-squared fit as determined by the $\chi^2$ values. (*a*) The rotation curve and (*b*) the position angle of the best ROTCUR fit are shown as a function of radius. The position angle here and in the text is measured from the North and increases in the counterclockwise direction.

**Figure 14** Velocity residuals of the CO field, taken as the observed velocity field minus the model velocity field based on the ROTCUR fits (see text). Positive velocity residuals are shown in the left panel, and negative velocity residuals are shown in the right panel. The velocity residuals are overlaid on the CO halftone. The contour levels are $\pm$ 10,20,30,40,50,60,70,80 km s$^{-1}$.

**Figure 15** Velocity fields of CO (left) and HCN (right) near the nucleus of NGC 1068. The contour interval is 50 km s$^{-1}$ over the range 1050–1250 km s$^{-1}$. The systemic velocity contour of 1130 km s$^{-1}$ is shown as a heavy line.

**Figure 16** Continuum map of nuclear emission from NGC 1068 at 85 GHz. The contour levels are $\pm$ 2,3,4,5 $\sigma$, where $\sigma = 8$ mJy beam$^{-1}$. The uniformly weighted beam is $4.2'' \times 3.4''$ and is shown in the lower left corner.

**Figure 17** The *Hubble Space Telescope* WFPC wideband image of NGC 1068. The center of the maps shown here is not aligned with the other maps shown in this paper. The *HST* image is shown as a halftone (left) and a contour map (right).

**Figure 18** The angular velocity $\Omega$ is shown along with the $\Omega \pm \kappa/2$ curves. The abscissa

shows the radius r measured in the plane of the galaxy. The corotation radius $r_L$ is taken to be 1.3 kpc (see text), so that the pattern speed of the bar $\Omega_b$ is 160 km s$^{-1}$ kpc$^{-1}$. The triangles are points computed from the CO data, while the open squares are from the HI data of Brinks et al. (1994).

**Figure 19** Ellipses that represent circles of constant radius in the plane of the galaxy are shown overlaid on the CO map of NGC 1068. The assumed position angle is taken to be 84° and the inclination angle is 40°. Ellipses are shown at r = 0.5, 1 and 1.5 kpc and are centered on the peak of the CO light distribution at (0″,-1″). The cross represents the kinematically determined center of rotation at (1.5″,-2.5″).

**Figure 20** Results from the LVG modeling of NGC 1068. (*a*) HCN results for gas at a kinetic temperature of 50 K. The solid contours show lines of constant $T_s$(1-0); from left, the levels are 1,5,10,20 K, with the measured value of 5 K shown in boldface. The ratio $T_s$(4-3)/$T_s$(1-0) are shown as dashed lines; from bottom to top, the levels are 0.3,0.7,1, with the measured 0.7 contour emphasized in boldface. (*b*) CO results for gas at a kinetic temperature of 50 K. The solid and dashed contours have the same relative meaning as in (*a*). The $T_s$(1-0) levels are 2,6,10,14,18 K, with the measured value of 10 K shown in boldface. The $T_s$(4-3)/$T_s$(1-0) levels are 1,2.5,5, with the measured value of 2.5 in boldface.